\documentclass[twocolumn,epjc3]{svjour3}

\usepackage[T1]{fontenc}
\usepackage[utf8]{inputenc}
\usepackage{color}
\usepackage{graphicx}
\usepackage{amsmath}
\usepackage{amsfonts}
\usepackage{mathdots}
\usepackage{mathrsfs} 
\usepackage{dsfont} 
\usepackage{stmaryrd}
\usepackage{enumerate}
\usepackage{hyperref}
\usepackage{appendix}
\usepackage{cancel}
\hypersetup{colorlinks=true,
            linkcolor=blue,
            citecolor=blue,
            filecolor=green,
            urlcolor=cyan 
}

\usepackage[numbers,sort&compress]{natbib}
\usepackage{scalefnt}
\usepackage{txfonts}
\usepackage{microtype}

\journalname{Eur. Phys. J. A}

\newcommand{\bra}[1]{\langle #1 \vert}
\newcommand{\ket}[1]{\vert #1 \rangle}
\newcommand{\scal}[2]{\langle #1 \vert #2 \rangle}
\newcommand{\elma}[3]{\bra{#1} #2 \ket{#3}}

\newcommand{\emax}{e_{\text{max}}}
\newcommand{\hbo}{\hbar \omega}

\newcommand{\basissize}{n_{\text{dim}}}

\graphicspath{{.}{figures/}}

\begin{document}

\title{Mean-field approximation on steroids: exact description of the deuteron}

\author{B.~Bally\thanksref{ad:esnt} \and 
A. Scalesi\thanksref{ad:saclay} \and 
V. Somà\thanksref{ad:saclay} \and
L. Zurek\thanksref{ad:bruyeres,ad:lmce} \and
T. Duguet\thanksref{ad:saclay,ad:kul}}
\date{Received: \today{} / Accepted: date}

\institute{
\label{ad:esnt}
ESNT, IRFU, CEA, Universit\'e Paris-Saclay, 91191 Gif-sur-Yvette, France 
\and
\label{ad:saclay}
IRFU, CEA, Universit\'e Paris-Saclay, 91191 Gif-sur-Yvette, France 
\and
\label{ad:bruyeres}
CEA, DAM, DIF, 91297 Arpajon, France
\and
\label{ad:lmce}
Université Paris-Saclay, CEA, Laboratoire Matière en Conditions Extrêmes, 91680 Bruyères-le-Châtel, France
\and
\label{ad:kul}
KU Leuven, Department of Physics and Astronomy, Instituut voor Kern- en Stralingsfysica, 3001 Leuven, Belgium 
}

\maketitle

\begin{abstract}
The present article demonstrates that the deuteron, i.e.\ the lightest bound nuclear system made of a single proton and a single neutron, can be accurately described within a mean-field-based framework. 

Although paradoxical at first glance, the deuteron ground-state binding energy, magnetic dipole moment, electric quad\-rupole moment and root-mean-square proton radius are indeed reproduced with sub-percent accuracy via a low-dimen\-sional linear combination of non-orthogonal Bogoliubov states, i.e.\ with a method whose numerical cost scales as $\basissize^4$, where $\basissize$ is the dimension of the basis of the one-body Hilbert space. By further putting the system into a harmonic trap, the neutron-proton scattering length and effective range in the ${}^{3}S_1$ channel are also accurately reproduced.

To achieve this task, (i) the inclusion of proton-neutron pairing through the mixing of proton and neutron single-particle states in the Bogoliubov transformation and (ii) the restoration of proton and neutron numbers before variation are shown to be mandatory ingredients. 

This unexpected result has implications regarding the most efficient way to capture necessary correlations as a function of nuclear mass and regarding the possibility to ensure order-by-order renormalizability of many-body calculations based on chiral or pionless effective field theories beyond light nuclei. In this context, the present study will be extended to $^{3}$H and $^{3,4}$He in the near future as well as to the leading order of pionless effective field theory. 
\end{abstract}

%=======================================================================
%=======================================================================
%=======================================================================
\section{Introduction}
\label{sec:intro}

\subsection{Context}

One of the main challenges of low-energy nuclear theory is the extension of the \emph{ab initio} description of nuclei to the heavy mass region. Based on Weinberg's power counting of chiral effective field theory ($\chi$EFT)~\cite{Weinberg1990a,Epelbaum:2008ga,Machleidt:2020vzm}, this requires to solve the $A$-body Schrödinger equation
\begin{equation}
\label{e:SE}
  H \, \ket{\Psi^{A}} = E^{A} \, \ket{\Psi^{A}} 
\end{equation}
to good enough accuracy from light to heavy systems based on a Hamiltonian $H$ constructed at a given order in the corresponding $\chi$EFT expansion. Employing pionless EFT ($\cancel{\pi}$EFT)\footnote{$\cancel{\pi}$EFT is characterized by a breakdown scale of the order of the pion mass~\cite{Ekstrom:2024dqr} that might eventually translate into an  intrinsic limited applicability domain with respect to the nuclear mass range.}~\cite{Hammer2020a} or alternative power countings of $\chi$EFT~\cite{yang_importance_2023, thim_perturbative_2024, machleidt_recent_2024}, the $A$-body Schrödinger equation must still be solved to good enough accuracy but now for the leading order (LO) part of the Hamiltonian only, the subleading terms being treated in perturbation with respect to the LO solution. 

In this context, solving Schrödinger's equation to ``good enough accuracy'' relates to two different, hopefully not exclusive, criteria, i.e.\ the accuracy must be high enough 
\begin{enumerate}
\item to meaningfully answer the empirical questions of interest,
\item to ensure the order-by-order renormalizability of the calculation such that results do not depend on an arbitrary cutoff used to regularize ultra-violet divergences in the theory~\cite{vanKolck:2020llt}.
\end{enumerate}
The accuracy needed to fulfill both criteria is indeed achieved in few-nucleon systems. This however relies on many-body methods tackling in full the exponential scaling with $A$ of the cost associated with solving Schrödinger's equation. Unfortunately, using such methods becomes quickly intractable for $A \geq 16$. 

Pushing \emph{ab initio} calculations to heavier/heavy nuclei makes necessary to circumvent this ``curse of dimensionality'' by systematically expressing the exact solution with respect to an appropriate zeroth-order state $ |\Phi \rangle$ according to~\cite{Hergert20a,Frosini22a}
\begin{equation}\label{e:waveop}
|\Psi^{A} \rangle = \Omega |\Phi \rangle \, ,
\end{equation}
where the wave operator $\Omega$ is to be expanded in a perturbative or non-perturbative fashion. Eventually, truncating this expansion to a given order makes (i) the cost of the method to be polynomial in $A$, (ii) the result systematically improvable towards the exact solution and (iii) possible to evaluate the error associated with omitted orders. 

The zeroth-order state $ |\Phi \rangle$ is meant to be obtained at mean-field-like cost, i.e.\ a cost that (naively) scales as $\basissize^4$ where $\basissize$ denotes the dimension of the (truncated) basis of the one-body Hilbert space. Today's common understanding is that strong {\it static} correlations emerging in open-shell systems can be efficiently captured through the use of versatile zeroth-order states exploiting the concept of symmetry breaking (and restoration) while possibly adding collective fluctuations~\cite{Yao18a,Yao20a,Yao:2022nv,Frosini22a,Frosini22b,Frosini22c,Bally2024a,Porro2024a,Porro2024b}. This can typically be achieved at $\basissize^4$ cost via a low-dimensional linear mixing of non-orthogonal mean-field product states within the frame of the projected generator coordinate method (PGCM). 
In the language used in this paper, the PGCM corresponds to a ``maximally boosted'' mean field and can be considered a variant of a mean field ``on steroids''.  

The remaining (weak) {\it dynamical} correlations are brought in consistently through the truncated expansion of the wave operator under the form of a large number of (hopefully) low-rank elementary, i.e.\ particle-hole or quasi-particle, excitations of the zeroth-order state\footnote{For a discussion on the nature of many-body correlations, i.e.\ weak/dynamical versus strong/static, the reader is referred to, e.g., Ref.~\cite{Hagen2022a}.}. The addition of dynamical correlations make the result more accurate but of course more costly, e.g.\ adding second- (third-) order perturbative corrections on top of a product state is achieved at an $\basissize^5$ ($\basissize^6$) cost~\cite{Tichai20a}.

Given the above, two key questions arise in relation to the two criteria listed above\footnote{The second question can only be addressed based on either $\cancel{\pi}$EFT or alternative power countings of $\chi$EFT given that Weinberg's power counting does not allow for an order-by-order renormalization~\cite{Hammer2020a}.}:
\begin{enumerate}
\item How do many-body correlations that must be accounted for to reach a ``good enough accuracy'' split between the zeroth-order state (i.e.\ static correlations) obtained at $\basissize^4$ cost and the wave operator expansion (i.e.\ dynamical correlations) captured at $\basissize^p$ cost with $p>4$? This question is critical in view of extending \emph{ab initio} calculations to heavy open-shell nuclei in an optimal way in the future~\cite{Scalesi2024a}.
\item How is the renormalizability affected by the fact that Schrödinger's equation cannot be solved exactly beyond few-nucleon systems~\cite{Drissi2020a}?
\end{enumerate}

\subsection{Sharing of many-body correlations}

The splitting between static and dynamical correlations depends on the nature of the reference state, i.e.\ on how much correlations can be captured while remaining at the (boosted) mean-field level. In Refs.~\cite{Frosini22a,Frosini22b,Frosini22c}, such a question was already touched upon via the use of a novel perturbative expansion (PGCM-PT) based on a PGCM zeroth-order state and a $\chi$EFT interaction evolved to low resolution scale via a similarity renormalization group (SRG) transformation~\cite{Bogner:2009bt,PhysRevLett.107.072501,PhysRevC.90.024325}. Applied to the mass $A=16$ -- $20$ region and using a small model space, it was demonstrated that~\cite{Frosini22b,Frosini22c}
\begin{itemize}
\item a PGCM, i.e.\ boosted mean-field, zeroth-order state efficiently captures static correlations arising in singly and doubly open-shell nuclei,
\item nuclei described via such a zeroth order state are significantly underbound, i.e.\ the binding energy is about $15\%$ above the exact result, 
\item missing correlations are consistently captured by low-order corrections from the wave operator expansion, i.e.\ adding the second-order correction via PGCM-PT(2), the binding energy is only about $1\%$ above the exact result,
\item while total energies are indeed underbound at the PGCM level, excitation energies of rotational excitations are seen to be accurately described, i.e.\ dynamical correlations added on top of the PGCM cancel to high accuracy between members of the ground-state rotational band.
\end{itemize}

The above demonstrates the high performance of a boost\-ed mean-field zeroth-order state even though accounting for dynamical correlations remains crucial. As a matter of fact, a less advanced version of the ``fully boosted'' mean-field PGCM state constituted by a single symmetry-breaking mean-field state is already an excellent starting point at and above the $A=20$ mass region even though the addition of dynamical correlations is obviously mandatory~\cite{Tichai18a,Tichai20a,Scalesi2024a}. 
Eventually, the performance of a well suited (boosted) mean-field state is expected to critically depend on the nuclear mass. In particular, a common belief is that the balance between static and dynamical correlations is largely unfavorable for the (boosted) mean-field zeroth-order state in very light systems such as the alpha particle or the deuteron. This expectation relies on the fact that the mean-field approximation is traditionally believed to be better justified for large systems where internal fluctuations are small compared to the average inter-particle interaction that defines
the mean field\footnote{This is supported by the observation that empirical energy density functional (EDF) models \cite{Bender03a} based on an effective mean-field approximation deliver a good coarse description of well-deformed heavy nuclei \cite{Bender06a,Dobaczewski15a}, although fine details or local variations with respect to the number of particles are not always reproduced.}. One observes, however, that the Bardeen-Cooper-Schrieffer approximation accounting for neutron-proton pairing ($np$BCS), i.e.\ the mean-field approximation breaking $U(1)$ global-gauge symmetry, seems to potentially contradict the above expectation when applied to homogeneous symmetric nuclear matter\footnote{Also, the EDF approach and its ``boosted'' extension is observed to work reasonably 
well when applied to rather light nuclei made of about ten particles or less \cite{Marevic2019a,Zhao2021a,Motoki2022a,Rong2023a}. The phenomenological nature of these calculations, however, makes difficult to disentangle the contributions of the mean-field approximation \emph{per se} and of the functionals that implicitly resum dynamical correlations.}. In the zero-density limit where the system reduces to a homogeneous gas of deuterons, the $np$BCS gap equation is indeed shown to reduce to the exact two-body Schrödinger equation for a (delocalized) deuteron and to deliver the corresponding binding energy~\cite{Baldo1995a}. 

\subsection{Renormalizability}

To which extent the boosted mean-field approximation can accurately describe the solution of the two-body Schrödinger equation is also highly relevant to address the question of the renormalizable character of the \emph{ab initio} theoretical scheme. In fact, e.g., the two-body part of the $\cancel{\pi}$EFT Hamiltonian at LO is renormalized in the two-body system by solving the corresponding Schrödinger equation exactly\footnote{$\cancel{\pi}$EFT indeed requires the re-summation of the $S$-wave two-body contact operators at LO to correctly account for the loosely bound  deuteron in the triplet channel and for the virtual state in the singlet channel. Next, the one-parameter three-nucleon contact interaction at LO, necessary to renormalize the three-nucleon system, is adjusted via essentially exact three-nucleon calculations~\cite{Schiavilla:2021dun}.}. 

The fact that the Schrödinger equation  can only be solved with ``good enough'' accuracy beyond light systems may compromise the initially built-in renormalization invariance of computed observables. For example, using a symmetry-conserving Hartree-Fock (HF) Slater determinant as zeroth-order state and complementing it with finite-order perturbative corrections leads to non-renormalizable LO calculations~\cite{Drissi2020a}. Identifying a zeroth-order state satisfying renormalization invariance at $\basissize^4$ cost might simplify the task of fulfilling it once dynamical corrections are included on top of it.

\subsection{Present work}

In view of the above, the goal of the present work is to characterize the performance of increasingly sophisticated versions of the boosted mean-field approximation to describe a real, i.e.\ finite-size and isolated, deuteron at $\basissize^4$ cost using a $\chi$EFT-based Hamiltonian~\cite{Entem2003a}. While presently concentrating on the lightest (non-trivial) nucleus over the nuclear chart, the study will be extended to three- and four-nucleon systems in the future. In a future publication we will also investigate if an appropriately designed boosted mean-field state can indeed satisfy renormalization invariance of $\cancel{\pi}$EFT at LO in the two-body system.

Thus, the present work focuses on the following questions:
\begin{itemize}
    \item Is the deuteron bound at the (boosted) mean-field level?
    \item If so, how accurately can the binding energy and other observables such as the electric quadrupole moment, the magnetic dipole moment and the proton root-mean-square radius be described?
    \item If such a description is indeed successful, what are the key characteristics of the boosted mean-field state necessary to describe the deuteron accurately? 
\end{itemize}
To address these questions, a hierarchy of variational methods centered around the ideas of symmetry-broken and \mbox{-}restored mean-field states~\cite{Duguet10a} is employed. 
Such concepts have proven to be very efficient in the study of atomic nuclei based on empirical EDFs \cite{Bender03a,Egido16a,Robledo18a} and other mesoscopic systems \cite{Sheikh21a}. More recently, and as already alluded to above, such boosted mean-field states have been promoted as versatile zeroth-order states in \emph{ab initio} calculations based on many-body expansion methods\footnote{The symmetry restoration, which will happen to be crucial in the present work, is not always performed in such studies. For examples of full realizations of these ideas in an \emph{ab initio} context, see Refs.~\cite{Yao20a,Hagen2022a,Frosini22a,Frosini22b,Frosini22c,Belley2024a}.} \cite{Soma13a,Soma2021a,Tichai18a,Novario2020a,Scalesi2024a}.

The article is organized as follows. 
Section \ref{sec:theory} introduces the various flavors of (boosted) mean-field approximations used in this work while Sec.~\ref{sec:results} discusses their ability to reproduce the structure of the deuteron.
Finally, conclusions and perspective are provided in Sec.~\ref{sec:conclu}.

%=======================================================================
%=======================================================================
%=======================================================================
\section{Theoretical framework}
\label{sec:theory}

%============
%============
\subsection{Hamiltonian and model space}
\label{sec:theory:ms}

The intrinsic Hamiltonian is written as 
\begin{equation}
  \label{eq:hint}
  H_\text{int} \equiv T_\text{tot} - T_\text{com} + V \, ,
\end{equation}
where $T_\text{tot}$ denotes the total kinetic energy, $T_\text{com}$ is the center-of-mass kinetic energy and $V$ is the $\chi$EFT-based EM500 two-body interaction~\cite{Entem2003a}. This Hamiltonian perfectly describes the key observables related to the $J^\pi_\epsilon = 1^+_1$ ground state of the deuteron~\cite{Entem2003a}.

The potential $V$ is further evolved via a unitary similarity renormalization group (SRG) transformation to a flow parameter value of 1.8 $\text{fm}^{-1}$ to produce the two-body part of the EM1.8/2.0 interaction introduced in Ref.~\cite{Hebeler11a}. Since the SRG evolution is fully unitary in the two-body Hilbert space, the converged results discussed in the present work are independent of the SRG transformation. Still, after SRG transformation the potential displays weaker coupling between low- and high-momentum two-body states such that the results converge faster with respect to the size of the one-body (and thus two-body) basis employed in the calculation. 

One-body (two-body) operators  written in second quantized form are presently expanded over the eigenbasis (tensor product of two eigenbases) of the one-body spherical harmonic oscillator (sHO) Hamiltonian. A one-body basis state, associated with creation/annihilation operators $\left\{ c_i^\dagger, c_i \right\}$, is characterized by a principal quantum number $n_i$, the orbital angular momentum $l_i$, the spin $s_i = 1/2$, the total angular momentum $j_i$ and its third component $m_{j_i}$, as well as by the isospin $t_i = 1/2$ and its third component $m_{t_i}$. Here, the convention $m_{t_i} = -1/2$ ($+1/2$) is employed for proton (neutron) single-particle states.

The model space, i.e.\ the truncated one-body basis, is characterized by two parameters: the harmonic oscillator energy spacing $\hbar \omega$, $\omega$ being the 
oscillator frequency, and the maximum energy quanta $\emax \equiv \max(2n +l)$ carried by a single-particle basis state.

%============
%============
\subsection{Schrödinger equation}
\label{sec:theory:se}

Rewritten with a more explicit account of all quantum numbers at play, the $A$-body Schrödinger equation [Eq.~\eqref{e:SE}] reads
\begin{equation}
  H_\text{int} \ket{\Psi^{Z N J M_J \pi}_\epsilon} = E_\epsilon^{Z N J \pi} \ket{\Psi^{Z N J M_J \pi}_\epsilon} \,  , \label{e:SE2} 
\end{equation}
where $\ket{\Psi^{Z N J M_J \pi}_\epsilon}$ is an eigenstate of $H_\text{int}$ with the eigenenergy $E_\epsilon^{Z N J \pi}$ characterized by several
symmetry quantum numbers: the proton number $Z$, the neutron number $N$, the total angular momentum $J$ and its third component $M_J$, as well as the parity $\pi$. The index $\epsilon=1,2,\ldots$ is used to label the various eigenstates sharing the same set of symmetry quantum numbers.

 \subsection{Center-of-mass contamination}
\label{sec:theory:com}

Exact solutions of Eq.~\eqref{e:SE2} exactly factorize into a center-of-mass part and an intrinsic part. This key property may be compromised by two practical considerations: 
\begin{enumerate}
\item the truncation of the $A$-body Hilbert space basis presently following from the tensor product of the truncated one-body Hilbert-space basis,
\item the approximate solving of Eq.~\eqref{e:SE2} over this truncated $A$-body Hilbert space.
\end{enumerate}

To tame down residual contaminations of the intrinsic wave function by components corresponding to excitations of the center of mass, it is possible to use
the Gloeckner-Lawson method~\cite{Gloeckner1974a} based on the modified Hamiltonian
\begin{equation}
\begin{split}
  \label{eq:lawson}
  H (\beta_L, \omega_L) &\equiv H_\text{int} + \beta_L  H_\text{com} (\omega_L)  , \\
   &\equiv H_\text{int} + \beta_L \left(T_\text{com} + \frac12 m A \omega_L^2 R_\text{com}^2 - \frac32 \hbar \omega_L \right) ,
  \end{split}
\end{equation}
where $m$ is the nucleon mass, $R_\text{com}$ is the center-of-mass position and $(\beta_L, \omega_L)$ are parameters. The additional term in Eq.~\eqref{eq:lawson} corresponds to a center-of-mass harmonic oscillator Hamiltonian that, for large enough values of $\beta_L$, suppresses contaminations from center-of-mass excitations by pushing them to higher energies. Similarly to what is done in 
No-Core Shell Model (NCSM) calculations \cite{Navratil2009a}, the frequency $\omega_L$ is presently taken to be the same as the frequency $\omega$ of the sHO one-body basis.

The Gloeckner-Lawson method is appropriate when the factorization is well realized in practice in spite of the two sources of difficulty listed above. While no formal justification exists, it was demonstrated that coupled cluster calculations with singles and doubles or the IMSRG method truncated at the two-body level display a satisfactory center-of-mass factorization in the ground-state wave-function using a center-of-mass HO Hamiltonian $H_\text{com}$ characterized by a frequency that is different from the one of the sHO basis~\cite{Hagen2009a,Hergert16a}. 

In the present work, the question of the factorization along with the application of the Gloeckner-Lawson method are addressed for the first time for the (boosted) mean-field approximation in the case of the deuteron. For the sake of simplicity,  $H (\beta_L, \omega)$ is denoted as $H$ in the remainder of the paper. The employed values of $\beta_L$ will be specified when discussing the numerical results. 

%============
%============
\subsection{Hartree-Fock mean-field approximation}
\label{sec:theory:hf}

The first, and simplest, variational scheme considered in this work is the HF method. The total energy of the system is minimized, 
\begin{equation}
  \delta \frac{\elma{\Phi}{H}{\Phi}}{{\scal{\Phi}{\Phi}}} = 0 \, ,
\end{equation}
within the manifold of Slater determinants
\begin{equation}
  \label{eq:productHF}
  \ket{\Phi} \equiv \prod_{i=1}^A a^\dagger_i \ket{0} ,
\end{equation}
where $\ket{0}$ is the bare vacuum and $\left\{ a_i^\dagger, a_i \right\}$ denote creation and annihilation operators obtained through the linear transformation
\begin{subequations}
\begin{align}
\label{eq:transfoHF}
&\begin{pmatrix}
    a  \\
    a^\dagger
\end{pmatrix}
= 
\begin{pmatrix}
    D^\dagger & 0  \\
    0       & D^T 
\end{pmatrix}
\begin{pmatrix}
    c  \\
    c^\dagger
\end{pmatrix} , \\
&D D^\dagger = D^\dagger  D = 1 .
\end{align}
\end{subequations}
Here, the matrix elements of $D$ represent the variational parameters to be determined.

The HF method is the standard embodiment of the mean-field approximation. Its most basic realization is the spherical HF (sHF) method within which the
reference state $\ket{\Phi}$ is constrained to be spherical with good angular momentum $J=0$. Such an approximation is only adapted to the description of doubly closed-shell nuclei as the product $\ket{\Phi}$ does not encapsulate any static correlations.
It is however possible to describe doubly open-shell nuclei while staying within the manifold of Slater determinants by allowing $\ket{\Phi}$ to break one or several symmetries of the nuclear Hamiltonian. Doing so, one effectively explores a larger variational space to capture mandatory static correlations, thus (usually) finding a Slater determinant with a total energy lower than the one of the sHF solution.
In this work, the deformed HF (dHF) approach allowing the solution to break parity and rotational invariances of the nuclear Hamiltonian is considered. 

While advantageous from an energetic point of view, the first ``boost'' associated with the breaking of spatial symmetries has the drawback that the state $\ket{\Phi}$ does not possess good parity $\pi$, total angular momentum $J$ and its third component $M_J$ anymore, i.e., the state decomposes as 
\begin{equation}
 \label{eq:decompoHF}
 \ket{\Phi} = \sum_{J M_J \pi} \sum_{\epsilon} c^{Z N J M_J \pi}_{\epsilon} \ket{\Theta^{Z N J M_J \pi}_{\epsilon}} \, ,
\end{equation}
where the coefficients $c^{Z N J M_J \pi}_{\epsilon}$ are complex numbers and $\ket{\Theta^{Z N J M_J \pi}_{\epsilon}}$ denotes (yet unspecified) orthogonal many-body wave functions carrying good symmetry quantum numbers.
One major limitation of states decomposing according to Eq.~\eqref{eq:decompoHF} relates to the evaluation of expectation values and transition matrix elements of 
low-energy observables. Indeed, such matrix elements typically depend sensitively on the symmetry quantum numbers of the initial and final states at play, e.g., on symmetry selection rules,
and it is clear from Eq.~\eqref{eq:decompoHF} that several states $\ket{\Theta^{Z N J M_J \pi}_{\epsilon}}$ with various quantum numbers may wrongly contribute to the final value.

%============
%============
\subsection{Hartree-Fock-Bogoliubov mean-field approximation}
\label{sec:theory:hfb}

The Hartree-Fock-Bogoliubov (HFB) method generalizes the HF approach to further include static pairing correlations within the mean-field picture at the price of breaking $U(1)$ global-gauge symmetry associated with particle number conservation. 
In this case, one still minimizes the total energy of the system, as in Eq.~\eqref{eq:productHF}, but exploring the larger variational space of Bogoliubov quasi-particle states, 
which are product states of the form
\begin{equation}
  \label{eq:productHFB}
  \ket{\Phi} \equiv \prod_{i} \beta_i \ket{0} ,
\end{equation}
with $\left\{ \beta_i^\dagger, \beta_i \right\}$ being quasi-particle creation and annihilation operators defined through the linear transformation
\begin{subequations}
\begin{align}
\label{eq:transfoHFB}
&\begin{pmatrix}
    \beta  \\
    \beta^\dagger
\end{pmatrix}
= 
\begin{pmatrix}
    U^\dagger & V^\dagger  \\
    V^T       & U^T 
\end{pmatrix}
\begin{pmatrix}
    c  \\
    c^\dagger
\end{pmatrix}
\equiv \mathcal{W} 
\begin{pmatrix}
    c  \\
    c^\dagger
\end{pmatrix}, \\
&\mathcal{W} \mathcal{W}^\dagger = \mathcal{W}^\dagger  \mathcal{W} = 1 .
\end{align}
\end{subequations}
No restriction is presently imposed on the Bogoliubov matrices $U$ and $V$, whose elements are the variational parameters of the problem, except for the fact they are kept real. In particular, the mixing of proton and neutron single-particle states necessary to open the neutron-proton pairing channel can be considered, 
which will happen to be crucial for the deuteron.
As all spatial symmetries are also allowed to break during the minimization, the corresponding scheme is labeled as d$np$HFB.

Because the Bogoliubov zeroth-order state breaks $U(1)$ global gauge symmetry, its decomposition is now more general and also includes a sum over proton and neutron numbers
\begin{equation}
 \label{eq:decompoHFB}
 \ket{\Phi} = \sum_{Z N J M_J \pi} \sum_{\epsilon} c^{Z N J M_J \pi}_{\epsilon} \ket{\Theta^{Z N J M_J \pi}_{\epsilon}} \, ,
\end{equation}
with all components having the same number parity $(-1)^{Z+N}$ equal to  either $+1$ or $-1$ \cite{Bally21a}. This also implies that the average proton and neutron numbers in $\ket{\Phi}$ have to be constrained to the
physical values throughout the minimization procedure \cite{RS80a}.

%============
%============
\subsection{Projection after variation (PAV)}
\label{sec:theory:pav}

Starting from a dHF or d$np$HFB state $\ket{\Phi}$, it is possible to
build more correlated zeroth-order states respecting symmetries of $H$. This further ``boosted'' mean-field approximation is obtained through the use of the symmetry quantum-number projection method \cite{Bally21a,Sheikh21a,Tanabe05a,RS80a} delivering the state
\begin{equation}
\label{eq:pav}
 \ket{\Theta^{Z N J M_J \pi}_{\epsilon}} \equiv  \sum_{K_J=-J}^J f^{Z N J M_J \pi}_{\epsilon K_J} P^Z P^N P^J_{M_J K_J} P^\pi \ket{\Phi} ,
\end{equation}
where the weights $f^{Z N J M_J \pi}_{\epsilon K_J}$ are complex numbers whereas $P^Z$, $P^N$, $P^J_{M_J K_J}$ and $P^\pi$ are projection operators onto 
good proton number, neutron number, angular momentum and parity, respectively. Their detailed expressions can be found elsewhere \cite{Bally21a,Sheikh21a,Tanabe05a,RS80a}. The variation within the dHF or d$np$HFB methods is performed first to obtain $\ket{\Phi}$ while the projection onto good symmetry
quantum numbers is performed afterwards. These ``projection after variation'' schemes are thus denoted as dHF+PAV and d$np$HFB+PAV.
Obviously, whenever the original state $\ket{\Phi}$ does not break a symmetry in the first place, the associated symmetry restoration is not required. For example, employing
Slater determinants makes $P^Z$ and $P^N$ superfluous and to be omitted.

The weights in Eq.~\eqref{eq:pav} are determined by applying the variational principle to the state $\ket{\Theta^{Z N J M_J \pi}_{\epsilon}}$, which can be understood as the diagonalization 
of the nuclear Hamiltonian within the subspace spanned by the states $\left\{ P^Z P^N P^J_{M_J K_J} P^\pi \ket{\Phi}, \right.$ 
$\left. \vphantom{P^N } K_J \in \llbracket -J, J \rrbracket \right\}$, 
with the label $\epsilon$ then denoting the order of the states in the resulting energy spectrum.  

From a quantum mechanical perspective, the projected states thus obtained have two main advantages. First, they carry good symmetry quantum numbers and transform appropriately under the corresponding symmetry rotations.  Second, they are linear superpositions of rotated product states and thus include additional static correlations associated with collective zero-energy, i.e.\ Goldstone-like, rotational modes.
 
\begin{figure}
    \centering
    \includegraphics[width=.99\linewidth]{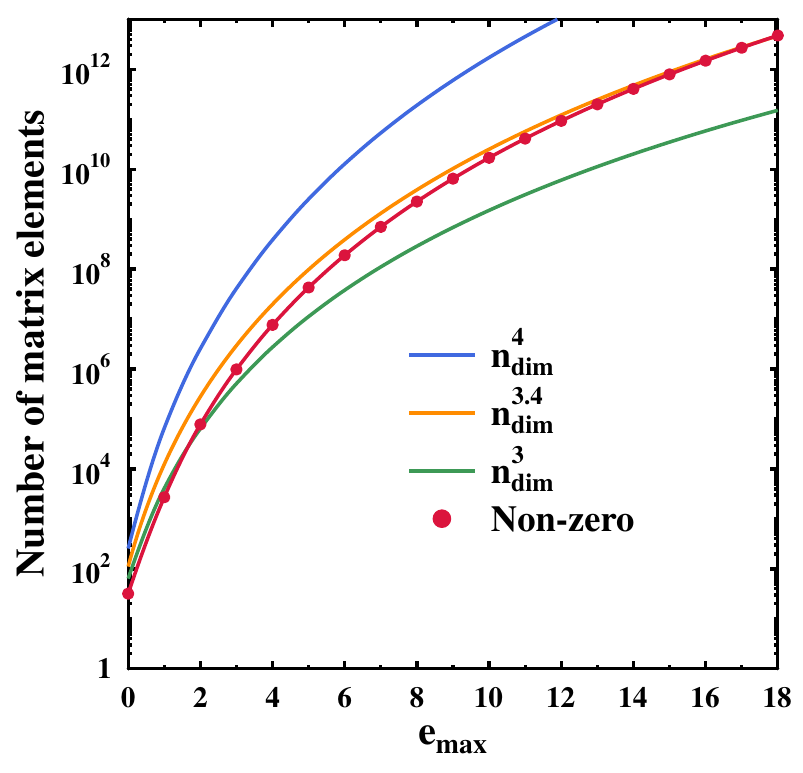} 
    \caption{Scaling of the number of two-body interaction matrix elements with the dimension 
    of the sHO one-body basis in m-scheme representation.
    Naive $\basissize^3$ (green line) and $\basissize^4$ (blue line) scalings are compared to the effective number of non-zero matrix elements resulting from the conservation of parity, projection of two-body angular momentum and two-body isospin on the $z$-axis (red circles). The CPU time of dHFB, projected dHFB or PGCM calculations is essentially proportional to the number of matrix elements entering the evaluation of Hamiltonian kernels between a pair of mean-field product states}
    \label{fig:scaling}
\end{figure}

From a practical point of view, the calculation of any expectation value or transition matrix element between projected states can be reduced to the calculation of 
matrix elements among all pairs of states belonging to the discrete\footnote{The integrals over the rotation angles appearing in the projection operators are discretized~\cite{Bally21a}.} set of rotated states forming the linear superposition mentioned above. Correspondingly, the computational scaling of the projection method retains the $\basissize^4$ scaling of the underlying mean-field method used to obtain $\ket{\Phi}$
times a prefactor\footnote{While often large, this prefactor is relatively independent of the model space and thus of the mass of the system under consideration~\cite{Bally21a}. Furthermore, methods are currently being developed to effectively reduce this prefactor~\cite{bofos24a}.}. In fact, and as illustrated in Fig.~\ref{fig:scaling}, the naive $\basissize^4$ scaling of projected HFB/PGCM calculations can be made closer to $\basissize^{3.4}$ over a large range of basis sizes by exploiting symmetries at hand when working in m-scheme with the sHO basis\footnote{Limiting here the illustration to two-body matrix elements, it is assumed that calculations beyond the two-body system would rely on a rank-reduction of the three-body interaction operator~\cite{Frosini21a} as is typically done in present-day \emph{ab initio} calculations beyond p-shell nuclei. A full account of three-body operator matrix elements would promote the naive scaling to $\basissize^6$ and the effective scaling closer to $\basissize^5$.}.

%============
%============
\subsection{Variation after particle-number projection (VAPNP)}
\label{sec:theory:vapnp}

An even more advanced, i.e.\ ``boosted'', mean-field-like approximation consists of performing the variational minimization in presence of the quantum-number projection, i.e.
of searching for the reference state $\ket{\Phi}$ yielding the lowest {\it projected} energy. Such a scheme is denoted as the ``variation after projection'' (VAP) approach.
In this work, VAP is performed only for the restoration of proton and neutron numbers such that the method is coined the d$np$VAPNP scheme.
Its master equation reads
\begin{equation}
  \label{eq:vap}
  \delta \frac{\elma{\Phi}{H P^Z P^N}{\Phi}}{{\scal{\Phi}{\Phi}}} = 0 \, ,
\end{equation}
where $\ket{\Phi}$ is a Bogoliubov quasi-particle state whose associated $(U,V)$ matrices still constitute the variational parameters of the problem. 

The minimized energy functional is different from the one at play in the d$np$HFB approach and corresponds to a state restricted to the Hilbert space associated with the desired proton and neutron numbers. Of course, the d$np$VAPNP calculation is more computationally demanding than the d$np$HFB+PAV as it requires to perform the particle-number projection at each iteration in the self-consistent solving of Eq.~\eqref{eq:vap}.

Finally, the d$np$VAPNP can be combined with a subsequent quantum-number projection on other symmetry quantum numbers to form the d$np$VAPNP\linebreak[2]+PAV method. 
This constitutes the most sophisticated approach presently considered and, for the sake of simple notations, will also be  referred to as the ``mean field on steroids'' (MFS) later on\footnote{As will become clear below, it is not necessary in the present study of the deuteron to invoke the most advanced MFS available, i.e.\ the PGCM.}.

%=======================================================================
%=======================================================================
%=======================================================================
\section{Results}
\label{sec:results}

The results shown in the present work have been obtained using the numerical suite TAURUS \cite{Bally2021a,Bally2024a} that permits to perform the sophisticated (boosted) mean-field calculations described above~\cite{Bally19a,Sanchez21a,Yao20a,Belley2024a,Giacalone2024a,Giacalone2024b}.

%============
%============
\subsection{Reference values}
\label{sec:results:ref}

In order to assess the accuracy of our description of the deuteron, results are compared to reference values obtained from NCSM calculations using the same interaction  for the ground-state  binding energy $E$, the electric quad\-ru\-pole moment $Q_s$, the magnetic dipole moment $\mu$ and the point-proton root-mean-square (rms) radius $r_{\text{rms},p}$~\cite{Miyagi2024b}, along with the values of the original fit for the scattering length $a_2$, and effective range $r_2$ ~\cite{Entem2003a}. Those reference values are reported in the second column of Tab.~\ref{tab:allval}. 

The operators $Q_s$ and  $\mu$ are presently limited to their one-body component~\cite{RS80a} whereas $r_{\text{rms},p}$ also includes the center-of-mass correction and thus contains one- and two-body contributions~\cite{Cipollone15a}. 
While consistently SRG-evolved operators~\cite{Miyagi2019a}, two-body currents \cite{Miyagi2024a} and relativistic corrections \cite{Entem2003a,Friar97a,Reinhard21a} would be required to perform a precise comparison to experimental data, their inclusion is irrelevant to the present study that focuses on the reproduction of the set of reference results. Of course, the NCSM reference calculations are performed using the same definition of the operators of interest.

\begin{table}[t]
\begin{center}
 \begin{tabular}{cccc}
   Quantity & Reference & MFS  & Relative error \\
   \hline 
   $J^\pi$ &  $1^+$     & $1^+$  & \\
   $E$ & $-2.2246$  & $-2.2225(12)$  & +0.09~\% \\
   $Q_s$  &  +0.2663  & +0.2643 & $-0.75$~\% \\
   $\mu$  & +0.8651  & +0.8649  & $-0.02$~\% \\
   $r_{\text{rms},p}$  & 1.982  & 1.975 & $-0.35$~\% \\
   $a_2$  & 5.417  & 5.44(3) & +0.42~\%\\
   $r_2$  & 1.752  & 1.71(5) & $-2.40$~\%
 \end{tabular}
\end{center}
\caption{\label{tab:allval} 
Observables associated with the deuteron ground state: spin-parity $J^\pi$, binding energy $E$ (MeV), electric quadrupole moment $Q_s$ ($e \text{fm}^2$), 
magnetic dipole moment $\mu$ ($\mu_N$), point-proton rms radius $r_{\text{rms},p}$ (fm), scattering length $a_2$ (fm) and effective range $r_2$ (fm).
Reference values are taken from Refs.~\cite{Miyagi2024b,Entem2003a} and are used to compute the relative errors of the MFS results obtained in this work
}
\end{table}

%============
%============

\subsection{Binding energy}
\label{sec:results:ener}

\begin{figure*}[t]
\centering
\includegraphics[width=0.99\linewidth]{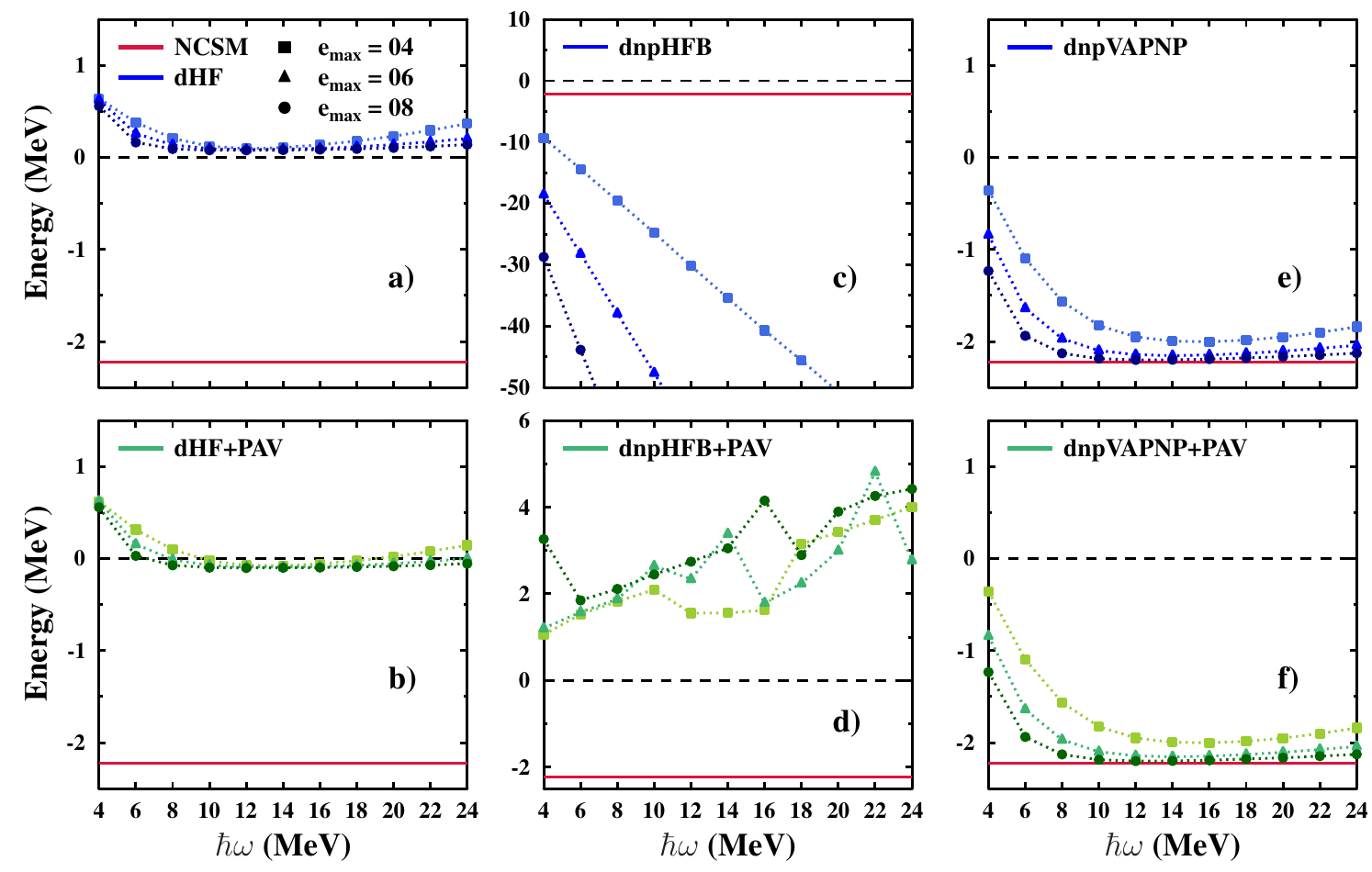}
\caption{Deuteron binding energy as a function of the oscillator spacing ($\hbar \omega$) for several sizes ($\emax$) of the sHO one-body basis. Results are provided for a) dHF,  b) dHF+PAV,  c) d$np$HFB,  d) d$np$HFB+PAV,  e) d$np$VAPNP and f) d$np$VAPNP+PAV (MFS) calculations.
The black dashed line represents the transition between a bound and an unbound nucleus.
The red line indicates the NCSM reference value reported in Tab.~\ref{tab:allval}}
\label{fig:energy_ingredients}
\end{figure*}

Let us first investigate the deuteron binding energy obtained with the pure intrinsic Hamiltonian, i.e., using $\beta_L = 0$ in Eq.~\eqref{eq:lawson}. The results obtained for $\beta_L \neq 0$ are reported and discussed in Sec.~\ref{sec:results:com}. 

Binding energies obtained from dHF, d$np$HFB and \linebreak[4] d$np$VAPNP calculations without and with subsequent PAV are displayed in Fig.~\ref{fig:energy_ingredients} as a function of $\hbar \omega$ for $\emax =4,6,8$. The NCSM result is also reported for reference. The dHF and d$np$HFB results relate to the plain unprojected energy whereas the d$np$VAPNP energy includes the sole projection on $(Z,N)=(1,1)$. Results labeled with PAV further include the projection on all symmetry quantum numbers $(Z,N,J,M_J,\pi)=(1,1,1,1,+)$. 

Starting with dHF results displayed in panel a), the deuter\-on is predicted unbound for all values of $\emax$ and $\hbar \omega$. Calculations are well converged with respect to the basis size for $\hbar \omega = 12$ MeV and give an energy of about $+77$ keV for $\emax = 8$. This result is in agreement with the folk knowledge stipulating that the description of the deuteron is qualitatively wrong at the plain HF level. Nevertheless, performing the PAV on top of the dHF calculations, the $1^+_1$ state is now bound with a total energy of about $-102$ keV for $\hbar \omega = 12$ MeV and 
$\emax = 8$ as visible from panel b). While the symmetry projection only provides a modest gain in energy, it radically changes the nature of the state, going from an unbound to a bound state. Still, the dHF+PAV energy is largely underbound compared to the NCSM reference value and, consequently, the method remains unsatisfying.

\begin{figure}[t]
\centering
\includegraphics[width=1.05\linewidth]{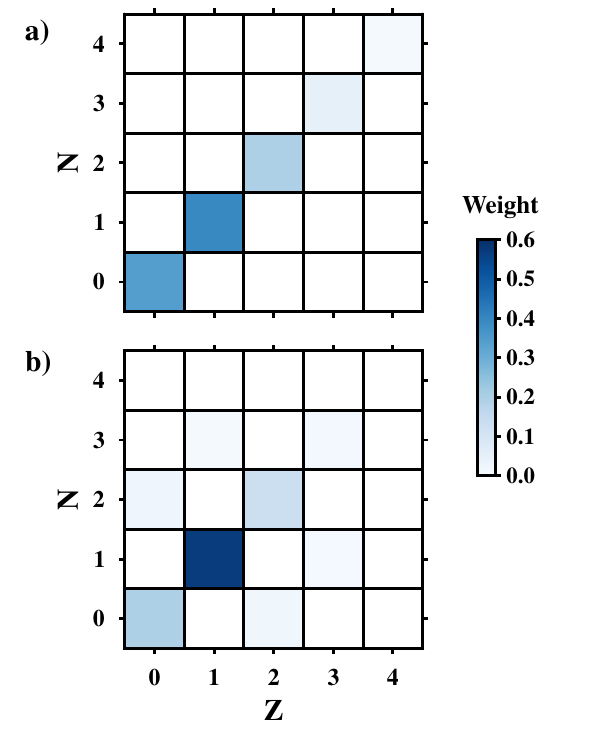}
\caption{Decomposition with respect to proton and neutron numbers of the Bogoliubov state originating from the a) d$np$HFB and b) d$np$VAPNP calculations for $\emax = 8$ and $\hbo = 12$ MeV}
\label{fig:NZ_decomposition}
\end{figure}

Boosting the HF mean-field approximation by allowing for neutron-proton pairing, the d$np$HFB energy shown in panel c) happens to be overbound by tens or hundreds of MeV. In fact, the results do not converge with respect to the basis size or the oscillator spacing.
This catastrophic behavior is due to the fact that the d$np$HFB method is a particle-number non-conserving theory. As a result, the expectation value of the energy receives contributions from components in the decomposition \eqref{eq:decompoHFB} with an incorrect number of particles, even if the mean values are constrained to $Z =N =1$. As shown in panel a) of Fig.~\ref{fig:NZ_decomposition}, the d$np$HFB solution obtained for $\emax = 8$ and $\hbo = 12$ MeV contains significant $N=Z$ components from the particle vacuum state ($N=Z=0$) through the alpha particle ($N=Z=2$) to $^8$Be ($N=Z=4$). As a result, the proton and neutron particle-number variances  are of the same order of magnitude as their mean value, e.g.\ 0.86 for $\emax = 8$ and $\hbo = 12$ MeV.
This feature is particularly problematic in the present case for three reasons. 
First, working with a soft two-body interaction, the repulsive contribution coming from the three-body sector needed to obtain a reasonable description of the neighboring nuclei appearing in the decomposition \eqref{eq:decompoHFB} is missing.
Second, the intrinsic Hamiltonian of Eq.~\eqref{eq:hint} is defined by fixing $A=2$ in the center-of-mass kinetic energy $T_\text{com}$, i.e., it does not take into account the large particle-number fluctuation presently characterizing the d$np$HFB
Bogoliubov state~\cite{Hergert09a}. Third, the binding energy of few-body systems changes rapidly with the number of nucleons, e.g., while the experimental binding energy of the deuteron is only of about 2 MeV, it raises to more than 28 MeV for ${}^{4}$He. As a result of these three features, the components with $N=Z>1$ anomalously drive the d$np$HFB energy to very negative values.

As shown in panel d) of Fig.~\ref{fig:energy_ingredients}, selecting the desired component of the d$np$HFB state with $Z=N=1$ through a subsequent PAV corrects for the catastrophic behavior but delivers an unbound deuteron with positive energies even larger than in the dHF calculations discussed above. Additionally, the energy exhibits an erratic pattern as a function of $\hbo$. Eventually, the variational process is too much affected by the presence of unphysical components and cannot be salvaged after the fact via the PAV.

The next step constituted by the d$np$VAPNP method delivers a drastic qualitative and quantitative improvement over the previous levels of (boosted) mean-field calculations. As visible from panel e) of Fig.~\ref{fig:energy_ingredients}, the deuteron does not only come out bound but the  d$np$VAPNP binding energy converges  with increasing values of $\emax$ towards the reference value provided by the NCSM calculation.
Even though d$np$HFB and d$np$VAPNP calculations explore the same manifold of Bogoliubov variational states, the superiority of the VAPNP minimization is obvious. Because the energy functional is associated with a many-body state residing in the Hilbert space associated with the correct proton ($Z=1$) and neutron ($N=1$) numbers, the calculation  is free from the problems mentioned above. As a matter of fact, and as can be seen from panel b) of  Fig.~\ref{fig:NZ_decomposition}, the decomposition of the underlying Bogoliubov state over eigenstates of the proton and neutron numbers is much more peaked around the $Z=N=1$ Hilbert space, the  $^8$Be component being completely suppressed\footnote{Interestingly, small components with $Z\neq N$ around $Z=N=1$ emerge at the same time.}. Effectively selecting the $Z=N=1$ component, the d$np$VAPNP eventually captures all correlations necessary to reproduce the deuteron binding energy.

The last step consists of adding the PAV on top of the d$np$VAPNP state, i.e.\ of further projecting on $(J,\pi)=(1,+)$. As seen in panel f) of Fig.~\ref{fig:energy_ingredients}, this subsequent PAV does not change significantly the results. More precisely, the energies are lowered by only a few keV, which is not visible on the scale used in the figure. This is easily explained by noticing that the d$np$VAPNP reference states obtained from the minimization procedure are already almost pure $J^\pi = 1^+$ states such that the impact of the PAV on the energy is marginal.

\begin{figure}[t!]
    \centering
    \includegraphics[width=.99\linewidth]{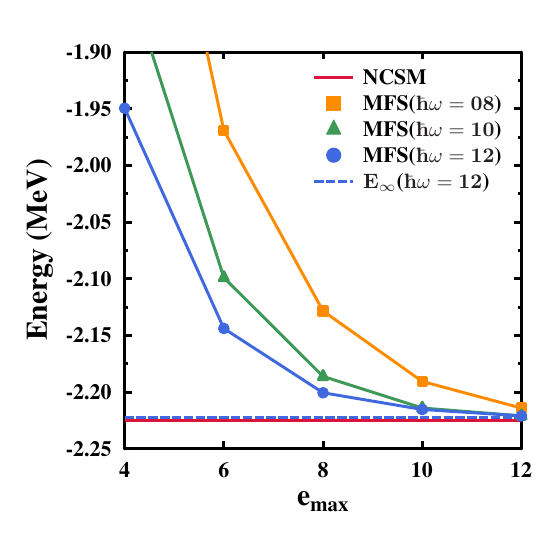} 
    \caption{Evolution of the energy from MFS calculations as a function of the basis size, $\emax$, for $\hbar \omega =$ 8 (yellow squares), 10 (green triangles) and 12 (blue circles) MeV. 
    The blue dashed line represents the extrapolation to infinite basis based on $\hbar \omega = 12$ MeV calculations.
    The red line indicates the NCSM reference value reported in Tab.~\ref{tab:allval}}
    \label{fig:energy_conv}
\end{figure}

Eventually, the d$np$VAPNP+PAV method, i.e.\ the most advanced boosted mean-field approximation under present consideration, delivers an accurate reproduction of the $J^\pi=1^+$ deuteron ground-state energy. The remainder of the article focuses on this ``mean field on steroids'' approach. To better gauge its accuracy, MFS computations have been carried out in larger model spaces, up to $\emax = 12$, for the three optimal values of $\hbo = 8, 10 \text{ and } 12$ MeV.  As can be seen from Fig.~\ref{fig:energy_conv}, the deuteron ground-state energy gently converges with respect to the basis size for all three values of $\hbo$ and the energy comes very close to the NCSM reference value at $\emax = 12$, e.g.\ it is equal to $E = -2.2209$\,MeV for $\hbo = 12$ MeV\footnote{The deuteron ground-state energy obtained from an exact diagonalization in relative coordinates using a $\emax = 12$ model space is equal to $E = -2.2228$\,MeV~\cite{Hagen2024P}, which is thus reproduced by the MFS calculation with an error of $1.9$\,keV.}. 

Given that the energies obtained for $\hbo = 12$ MeV display the smoothest convergence, they are used to perform an infinite basis-size extrapolation by fitting the results with a function of the form: $a \exp(-b \emax) + E_\infty$ \cite{Furnstahl2015a}, where $a, b$ and $E_\infty$ are parameters. The extrapolated value $E_\infty = -2.2225(12)$ MeV is marked by a blue dashed line in Fig.~\ref{fig:energy_conv} and reported in Tab.~\ref{tab:allval}. The figures in parentheses represent the uncertainty associated with a $95~\%$ confidence level assuming a normal distribution of the error with a variance estimated from the determination of the covariance matrix. The extrapolated mean value is in spectacular agreement with the NCSM reference value of $-2.2246$\,MeV with an error of only about 2 keV, which represents a relative error of $+0.09~\%$. 
It can thus be concluded that the MFS calculations are essentially exact\footnote{Performing $\emax = 14$ calculations and including them in the fit would probably help further reduce the error. Still, a point is reached where small numerical differences, e.g., in the generation of matrix elements of the interaction, may become the limiting factor in the reproduction of the NCSM reference value.} as far as the energy is concerned\footnote{Given that the SRG transformation is strictly unitary in the two-body system, observables would be unchanged by construction when using less-evolved versions of the EM500 interaction when solving the two-body Schrödinger equation exactly. As explained in Sec.~\ref{sec:results:analysis}, the ``mean field on steroids'' ansatz does indeed incorporate the necessary physics to converge towards the exact two-body solution such that this statement applies to it. Of course, such a convergence would, as with any other appropriate many-body technique, be slower with respect to the basis size when using such less-evolved versions of the EM500 interaction.}. 

\subsection{Analysis}
\label{sec:results:analysis}

The set of results presented above demonstrates that the necessary and sufficient conditions for a ``boosted'' mean-field approximation to deliver the deuteron binding energy essentially exactly\footnote{Employing an explicit expansion on top of a mean-field zeroth-order state, the deuteron can be solved exactly without resorting to these two ingredients. For example, a coupled cluster calculation performed on top of the sHF reference state at the singles and doubles (CCSD) level delivers an exact solution of the deuteron at $n_p^4n_h^2\sim \basissize^4$ cost, where $n_p$ ($n_h=2$) denotes the number of unoccupied (occupied) orbitals in the reference Slater determinant. The binding energy obtained with CCSD at $\emax=12$ is $-2.2211$\,MeV~\cite{Hagen2024P}, which is to be compared to $-2.2209$\,MeV computed with the MFS.} are
\begin{enumerate}
\item to allow for proton-neutron pairing,
\item to restore good neutron and proton numbers {\it before} variation.
\end{enumerate}

Solving the two-body problem exactly corresponds to summing to all orders the set of so-called particle-particle (particle-particle and hole-hole) ladder diagrams with respect to the particle vacuum (to a Slater determinant with two particles) making up the (in-medium) T matrix. It happens that the BCS gap equation extracts the pole of the (in-medium) T matrix associated with the occurrence of a two-body bound state in the nuclear medium~\cite{balian62a}, i.e.\ the Cooper pair. As a matter of fact, the $np$BCS gap equation computed for symmetric nuclear matter was shown to reduce in the zero-density limit characterizing a homogeneous gas of deuterons to the exact two-body Schrödinger equation for a (delocalized) deuteron~\cite{Baldo1995a}. This important result clearly underlines the necessity, when restricting oneself to the class of mean-field product states, to tackle neutron-proton pairing correlations in order to describe the deuteron accurately.  

However, capturing proton-neutron correlations is accomplished at the price of breaking $U(1)$ symmetry associated with the conservation of proton and neutron numbers. While the impact of such a symmetry breaking is marginal for the infinite homogeneous system of deuterons discussed in Ref.~\cite{Baldo1995a}, the d$np$HFB results discussed above demonstrate that it leads to a catastrophic behavior in a small system such as an isolated deuteron due to the fact that the particle-number fluctuations are of the same order as the particle number itself. This is the reason why the particle-number projection, performed before variation, is eventually mandatory to benefit from neutron-proton pairing correlations while properly accounting for the small size of the deuteron.

%============
%============
\subsection{Spectroscopic observables}
\label{sec:results:mu}

\begin{figure}[t!]
    \centering
    \includegraphics[width=.90\linewidth]{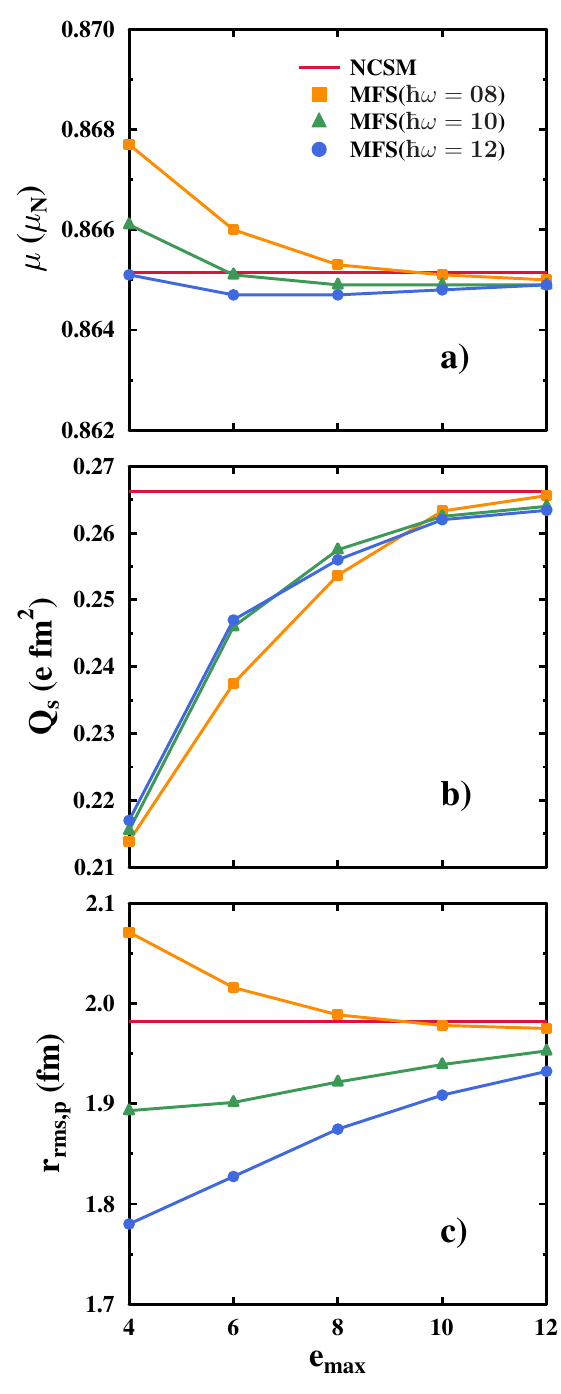} 
    \caption{Evolution of spectroscopic observables obtained from MFS calculations as a function of the basis size, $\emax$, for $\hbar \omega =$ 8 (yellow squares), 10 (green triangles) and 12 (blue circles) MeV. Panel a): magnetic dipole moment, panel b): electric quadrupole moment, panel c): point-proton rms radius. In each case, the red line indicates the NCSM reference value reported in Tab.~\ref{tab:allval}}
     \label{fig:muqsrp}
\end{figure}

An accurate description of the deuteron ground state $\ket{\Psi^{1111+}_1}$ not only requires the reproduction of the binding energy but also of spectroscopic observables such as the magnetic dipole, the electric quadrupole moments and the point-proton rms radius, respectively computed  as 
\begin{subequations}
\label{defoperator}
\begin{align}
  \mu &= \elma{\Psi^{1111+}_1}{g_l l + g_s s}{\Psi^{1111+}_1} \, , \\
   Q_s &= \sqrt{\frac{16\pi}{5}} \, \elma{\Psi^{1111+}_1}{r^2_p Y_{20}}{\Psi^{1111+}_1} \, , \\
   r_{\text{rms},p} &=  \sqrt{\elma{\Psi^{1111+}_1}{r^2_{p+\text{com}}}{\Psi^{1111+}_1}} \, .
\end{align}
\end{subequations}
In Eqs.~\eqref{defoperator}, $g_l$ denotes the orbital g-factor, $l$ the orbital angular momentum operator, $g_s$ the spin g-factor, $s$ the spin operator, $r_{p(+\text{com})}$ the position operator for protons (corrected for the center of mass) and $Y_{20}$ the spherical harmonic of degree 2 and order 0. The bare values of the g-factors obtained from the CODATA compilation \cite{CODATA2022} are used here.

The MFS results obtained for $\hbo = 8,10,12$ MeV are displayed in Fig.~\ref{fig:muqsrp} as a function of $\emax$ and are compared to the NCSM reference values. The best MFS values (see below) are also reported in Tab.~\ref{tab:allval}. For reasons explained in Sec.~\ref{sec:results:com}, these calculations  employ $\beta_L = 1$.

\begin{figure}[t!]
    \centering
    \includegraphics[width=.99\linewidth]{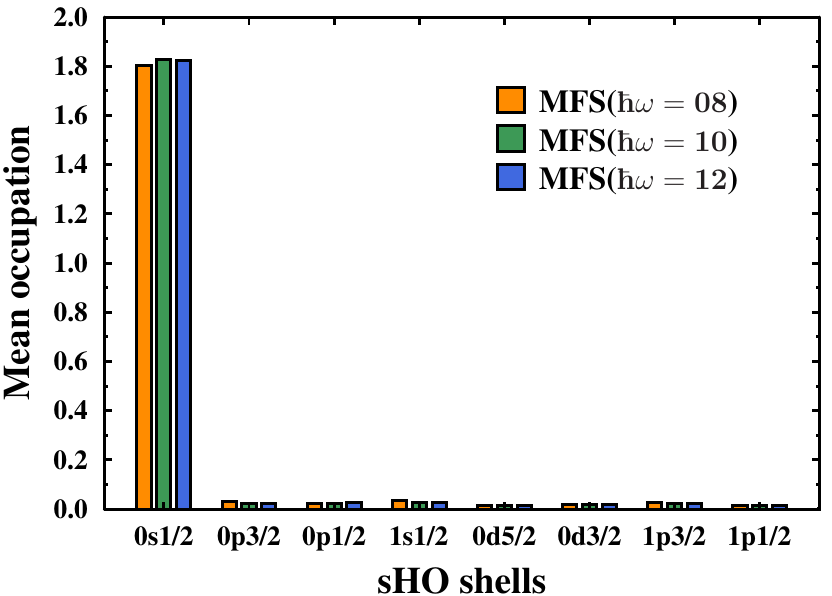} 
    \caption{Mean occupation of the lowest-lying one-body sHO shells extracted the MFS calculations of the deuteron ground state with $\emax = 12$ and $\hbar \omega =$ 8 (yellow bars), 10 (green bars) and 12 (blue bars) MeV}
     \label{fig:occup}
\end{figure}

The magnetic dipole moment shown in panel a) displays a gentle convergence as function of $\emax$ with the values obtained with the three $\hbo$ values giving the same result at a $10^{-4}$ level for $\emax = 12$. Their average value $\mu = +0.8649 \, \mu_N$ agrees with the NCSM reference value $+0.8651 \, \mu_N$ with a relative error of $-0.02$~\%. Interestingly, while the dHF+PAV method largely underestimates the deuteron binding energy, it gives a reasonable value of $+0.878 \, \mu_N$ for the magnetic moment. This is probably due to the fact that the magnetic moment is largely determined by 
contributions coming from the 0s1/2 single-particle states, which are the most occupied single-particle states in both dHF+PAV and MFS calculations, as is illustrated in Fig.~\ref{fig:occup} for MFS.

The electric quadrupole moment shown in panel b) of Fig.~\ref{fig:muqsrp} also converges smoothly as a function of $\emax$. The remaining dependence on $\hbo$ at $\emax=12$ is however slightly larger than for $\mu$. In absence of a controlled extrapolation method, the average of the three values $Q_s = +0.2643 \, e \text{fm}^2$ is considered and shown to agree with the NCSM reference value $Q_s = +0.2663 \, e \text{fm}^2$ with a relative error of $-0.75$~\%.
The small electric quadrupole moment of the deuteron is usually related to a (non-observable) small relative $D$-wave component of its wave function. 
The (non-observable) single-particle occupation numbers extracted from the MFS wave-function and displayed in Fig.~\ref{fig:occup} underline that the p- and d-shells  
have small but non-vanishing occupations, which is indeed consistent with this interpretation in the present calculation.

Finally, panel c) of Fig.~\ref{fig:muqsrp} reports the point-proton rms radius. While  MFS results obtained for the three $\hbo$ values show more variations than for $\mu$ and $Q_s$, the set associated with $\hbo = 8$ MeV displays the best convergence of the three. This is consistent with previous empirical observations that radii converge best for a smaller value of $\hbo$ than the one being optimal for the energy~\cite{Bogner2008a,Wolfgruber2024a}. This seems to be particularly true for the deuteron that is a halo nucleus with a very large point-proton (or charge) rms radius \cite{Hammer2023a}.
For $\emax = 12$ and $\hbo = 8$ MeV, the MFS value $r_{\text{rms},p} = 1.975$\,fm compares very well with the NCSM reference value $r_{\text{rms},p} = 1.983$\,fm. Indeed, it amounts to a relative error of only $-0.35$~\%. Quite similarly, CCSD delivers $r_{\text{rms},p} = 1,973$\,fm~\cite{Hagen2024P}.

In conclusion, the MFS is able to describe all spectroscopic observables associated with the $1^+_1$ deuteron ground-state with a sub-percent accuracy. 

%============
%============
\subsection{$np$ scattering in the ${}^{3}S_1$ channel}
\label{sec:scattering}

From the perspective of nuclear scattering theory, the deuter\-on represents the bound state solution of $np$ scattering in the 
${}^{3}S_1$ partial wave. One can thus expect that to an accurate description of the deuteron ground state corresponds an equally good description of the low-energy scattering properties in this channel, i.e.\ scattering length $a_2$ and the effective range $r_2$.  

In seminal work performed in the context of Lattice Quantum Chromodynamics calculations, the box-size dependence of the energy levels of two interacting particles in a finite box was shown to be related to their elastic scattering phase shifts in the infinite volume~\cite{Luscher1986a}.
Similarly, the scattering properties of two interacting particles can be accessed by trapping them in a (one-body) harmonic potential of varying frequency and extracting the eigenvalues of the corresponding static Schrödinger equation~\cite{Busch1998a,Stetcu2010a}.
Correspondingly, a neutron and a proton are now trapped in a harmonic trap with frequency $\omega_t$ according to the Hamiltonian
 \begin{equation}
 \label{eq:trap}
    H_{\text{trap}} (\omega_t) = H(0,\omega) +  \frac12 m \omega_t^2 \sum_{i=n,p} r^2_i \, ,
 \end{equation}
with $r_n$ ($r_p$) the neutron (proton) position operator. The $J^\pi=1^+$ ground-state energy of $H_{\text{trap}} (\omega_t)$  obtained from the MFS calculation is denoted as  $E_t$. 
For reasons explained in Sec.~\ref{sec:results:com}, the Gloeckner-Lawson term is omitted here, i.e.\ $\beta_L = 0$.

The so-called ``BERW formula'' \cite{Stetcu2010a,Guo2021a,Zhang2024a,Zhang2024b} permits to relate $E_t$ to the phase shifts $\delta_L (k)$ for a given partial wave with relative angular momentum $L$ at linear momentum $k = \sqrt{2(\mu_m c^2) E_t}/\hbar c$, where $\mu_m = m/2$ is the reduced mass. Focusing on the spherical wave $L = 0$, the relation reads
\begin{equation}
\label{eq:busch1}
-2 \frac{\sqrt{(\mu_m c^2) \hbar \omega_t}}{\hbar c} \frac{\Gamma(\frac34 - \frac{E_t}{2\hbar \omega_t})}{\Gamma(\frac14 - \frac{E_t}{2\hbar \omega_t})} 
= k \cot(\delta_0[k]) \, .
\end{equation}
The effective range expansion at second order [ERE(2)]~\cite{bethe49a}, which is justified at low energies, allows one to re-express the right-hand side of Eq.~\eqref{eq:busch1} as 
\begin{equation}
\label{eq:ere2}
 k \cot(\delta_0[k]) = -\frac{1}{a_2} + \frac12 r_2 k^2 + o(k^2) \, .
\end{equation}
In fact, one needs here to rather invoke the continuation of the ERE to imaginary momentum that relates the energy of the shallow two-body bound state (presently in the trap) associated with the pole in the S-matrix to the scattering length and effective range according to~\cite{Kievsky:2021ghz}
\begin{equation}
\label{eq:busch2}
-2 \frac{\sqrt{(\mu_m c^2) \hbar \omega_t}}{\hbar c} \frac{\Gamma(\frac34 - \frac{E_t}{2\hbar \omega_t})}{\Gamma(\frac14 - \frac{E_t}{2\hbar \omega_t})} 
= -\frac{1}{a_2} + \frac{(\mu_m c^2) r_2}{(\hbar c)^2} E_t \, .
\end{equation}
To estimate the values of $a_2$ and $r_2$, the strategy consists in solving the Schrödinger equation associated with $H_{\text{trap}} (\omega_t)$ for various values of $\omega_t$ and in performing a least-square linear regression of the ratio on the left-hand side (lhs) of Eq.~\eqref{eq:busch2} as a function of $E_t$. In order to satisfy the correct asymptotic behavior, the reduced oscillator length of the trap, $b_t\equiv \sqrt{\hbar/(\mu_m\omega_t)}$, must be taken larger than the range of the interaction. Using the reference value of $a_2=5.417$\,fm in the ${}^{3}S_1$ channel as a conservative estimate of this range, it amounts to considering traps characterized by $\hbar \omega_t \leq \frac{\hbar^2}{\mu_m a_2^2} \approx 3$ MeV.

\begin{figure}[t!]
    \centering
    \includegraphics[width=.99\linewidth]{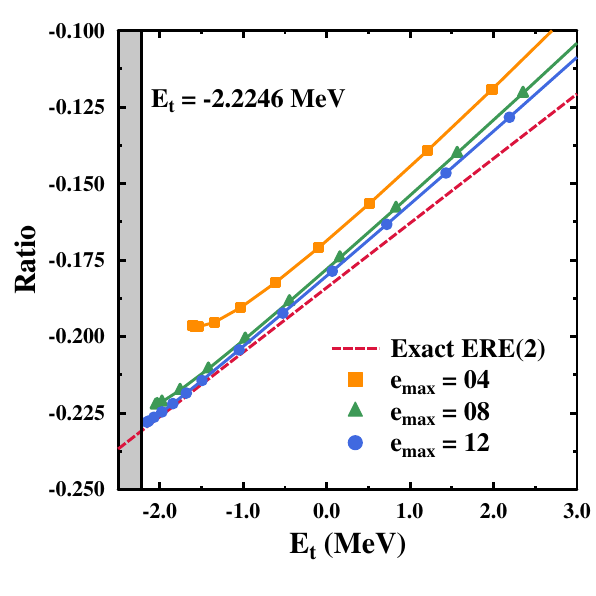} 
    \caption{Value of the ratio on the lhs of Eq.~\eqref{eq:busch2} as a function of the energy in the trap $E_t$. Full lines represent MFS calculations with different values for $\emax$ for $\hbar \omega = 30$ MeV. The dashed line represents the ERE at second order using exact values for $a_2$ and $r_2$}
     \label{fig:trap_pav}
\end{figure}

The ratio on the lhs of Eq.~\eqref{eq:busch2} is displayed in Fig.~\ref{fig:trap_pav} as a function of $E_t$ for different values of $\emax$ and $\hbar \omega = 30$ MeV. This particular value of  $\hbar \omega$ is motivated by the fact that the convergence of the scattering amplitude requires a large value of the ultraviolet cutoff associated with the sHO basis as demonstrated in Ref.~\cite{Stetcu2010a}.
As can be seen, the values of the ratio converge with increasing values of $\emax$. Considering $\emax=12$ and low values of $E_t$, the ratio is very close to the ideal linear behavior of the ERE at second order computed with the reference values $a_2=5.417$\,fm and $r_2=1.752$\,fm represented by the red dashed line. By contrast, as the value of $E_t$ increases, the curve departs from the ERE(2) curve. This is the result of two different effects. First, as the energy $E_t$ increases, higher-order terms in the ERE departing from the linear behavior become more important. Second, this energy region corresponds to larger values of $\hbar \omega_t$ for which the assumptions used to derive the BERW fomula are not satisfied anymore.

Finally, taking $\emax=12$, $\hbar \omega = 30$ MeV and focusing on the region $\hbar \omega_t \leq 3.0 $ MeV ($E_t \leq -1.5$ MeV), the linear regression delivers $a_2 = 5.44(3)$ fm and $r_2 = 1.71(5)$ fm. The figures in parentheses represent the uncertainty estimate on the fit parameters assuming a Student's t distribution with a 95~\% confidence level.
The obtained values are in excellent agreement with the reference ones and demonstrate once again the high accuracy of the MFS approximation. While the mean value of $r_2$ is less accurate than the results obtained for other observables, the exact value does fall within the confidence interval and it is expected that pushing the calculation to $\emax=14$ would further improve the result.

%============
%============
\subsection{Treatment of the center of mass}
\label{sec:results:com}

\begin{figure}[t!]
    \centering
    \includegraphics[width=.99\linewidth]{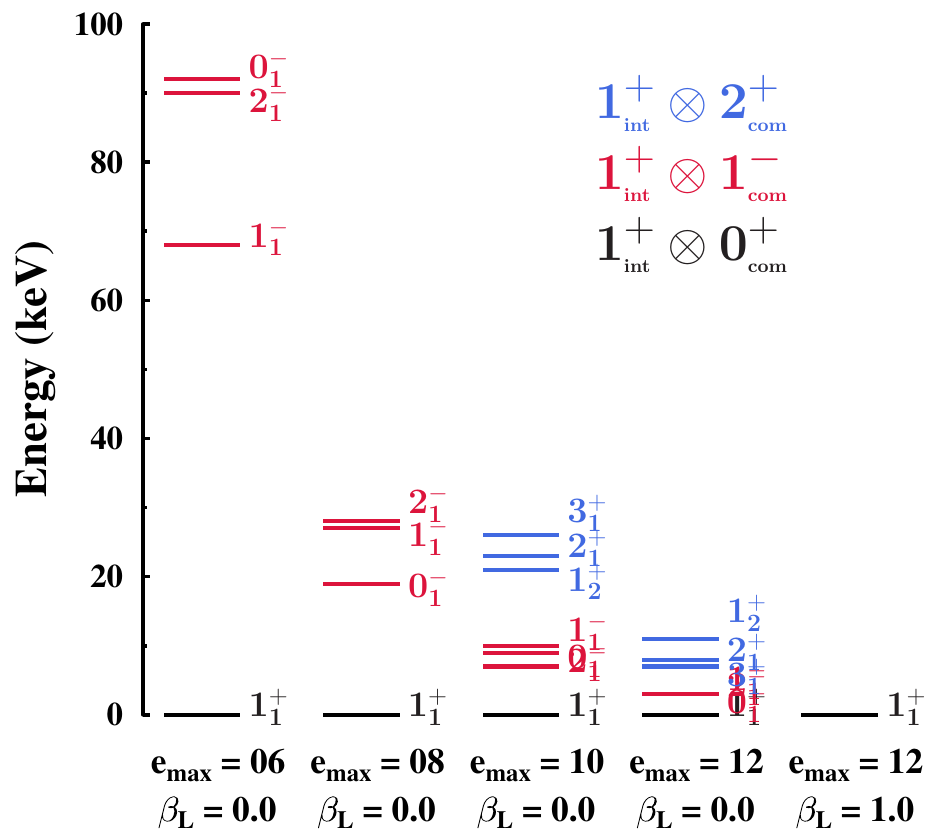} 
    \caption{Energy spectrum of the deuteron, up to 100 keV, from MFS calculations with different values of $\emax$ and $\beta_L$. The oscillator spacing is fixed to  $\hbo = 12$ MeV. Different colors are used to separate the states according to their dominant center-of-mass component}
    \label{fig:energy_spectrum}
\end{figure}

Working with the intrinsic Hamiltonian defined in Eq.~\eqref{eq:hint} can lead to the presence of ``spurious'' states in the intrinsic energy spectrum of the deuteron corresponding to copies of the spectrum of $H_\text{int}$ built on top of each 
center-of-mass eigenstate. This is illustrated in Fig.~\ref{fig:energy_spectrum} where the energy spectrum of the MFS calculation is shown up to 100 keV for $\hbo =  12$ MeV and various values of $\emax$ and $\beta_L$. 
Considering the pure intrinsic Hamiltonian ($\beta_L = 0$) the spectrum contains spurious states associated with the coupling of the intrinsic $1^+$ ground state to the center-of-mass eigenstates with $J^\pi = 0^+, 1^-, 2^+$. 
As the basis size increases, the states become more and more bunched together. This behavior is logical as one expects all states to be degenerate in an infinite basis.

The main difficulty relates to the fact that as the states come closer and closer in energy a spurious mixing can occur such that the projected states are possibly not pure center-of-mass eigenstates. For example, while the $1^+_1$ MFS state obtained for $\emax = 12$ and $\beta_L = 0$ can be safely assumed to mainly relate to the $0^+$ center-of-mass eigenstate, it still is partly contaminated by the coupling of the intrinsic $1^+$ state to the $2^+$ center-of-mass eigenstate. Also, it appears that small projected components with various values of $K$ appear in the decomposition \eqref{eq:decompoHFB} of the reference states. Unfortunately, it is not possible to address this problem by simply cutting off these components as it would imply hand-picking specific, and possibly large, cutoff values for each single calculation, rendering the results highly cutoff dependent. 

To remove the spurious center-of-mass contaminations, the modified Hamiltonian $H(\beta_L, \omega)$ introduced in Eq.~\eqref{eq:lawson} is used with $\beta_L > 0$. As seen in Fig.~\ref{fig:energy_spectrum}, the deuteron spectrum obtained for
$\emax = 12$ and $\beta_L = 1.0$ then only consists of the $1^+_1$ ground state. 
The first (and only) spurious excitation corresponding to a projected component with a very small projected overlap that could safely be removed
appears several MeV above the particle-emission threshold.
Furthermore, the expectation value of the center-of-mass Hamiltonian in the underlying reference state $\ket{\Phi (1.0)}$ obtained 
after solving Eq.~\eqref{eq:vap} is $\elma{\Phi (1.0)}{H_\text{com}(\omega)P^Z P^N}{\Phi(1.0)} = 0.001$ MeV,
which is to be compared to $\bra{\Phi (0.0)} H_\text{com}(\omega)P^Z P^N$ $\ket{\Phi(0.0)} = 1.227$ MeV obtained with $\beta_L = 0$. 
In conclusion, setting $\beta_L = 1$ removes any center-of-mass excitation and delivers a $1^+_1$ ground state purely built on top of the $0^+$ center-of-mass eigenstate.
Interestingly, the fact that the Gloeckner-Lawson method works as intended can be interpreted as an indirect proof that our method indeed  converges towards the exact solution with increasing basis size.

Nevertheless, a side effect of applying the Gloeckner-Lawson method is that for smaller values of $\emax$ or other choices of $\hbo$ for which the energy is not as well converged, the value of the energy is noticeably impacted.
For example, considering $\hbo = 12$ MeV and going from $\beta_L = 0$ to $\beta_L = 1$, the binding energy increases by about 91 keV for $\emax = 4$,
27 keV for $\emax = 6$, 7 keV for $\emax = 8$, 940 eV for $\emax = 10$ and 610 eV for $\emax = 12$. For other values of $\hbo$, the situation is slightly worse.
By contrast, the presence of spurious components in the spectrum did not seem to have an effect on the ground-state energy, which is why $\beta_L=0$ was used in Secs.~\ref{sec:results:ener} and~\ref{sec:scattering}.  For example, using $\emax = 12$ and $\hbo = 12$ MeV for which the center-of-mass contamination is maximal (see Fig.~\ref{fig:energy_spectrum}), the ground-state energy changes by less than 1 keV when removing the spurious states. In any case, using $\beta_L = 1$ for the MFS calculations based on $\hbo = 12$ MeV leads to an extrapolated ground-state energy of $E_\infty = -2.2234(12)$, which is also in excellent agreement with the value reported in Tab.~\ref{tab:allval}.

On the other hand, the calculation of the spectroscopic observables does require the removal of spurious states, hence the use of $\beta_L = 1$ in Sec.~\ref{sec:results:mu}. 
While in fact not really necessary for the magnetic dipole moment, it is mandatory to safely compute the electric quadrupole moment and the point-proton rms radius\footnote{
It is observed that the projection after variation on $(J,\pi)=(1,+)$ was only necessary for $\beta_L=0$ to remove the small artificial components due to the center-of-mass contamination. For a value of $\beta_L$ large enough, the d$np$VAPNP is seen to already deliver a pure $J^\pi=1^+$ state and the PAV can be avoided altogether.}. Indeed, the values of 
$Q_s$ and $r_{\text{rms},p}$ are fully unreliable for $\beta_L = 0$ as they exhibit an extreme sensitivity to the mixing of different $K$-components.    
The reason behind this difference is probably that, while the magnetic dipole moment is principally determined by the underlying occupation of the 0s1/2 shell, $Q_s$ and $r_{\text{rms},p}$ have a spatial character that is  highly sensitive to the contamination coming from center-of-mass excited states. Of course, the final values were checked to be robust against a reasonable variation of $\beta_L$ by performing additional calculations with $\beta_L = 0.5$ and $2.0$.

%=======================================================================
%=======================================================================
%======================================================================= 
\section{Conclusions}
\label{sec:conclu}

This article investigated the possibility to accurately describe the deuteron within a mean-field-based framework, i.e.\ without the need to further add missing dynamical correlations via an expansion method built on top of such a mean-field-based zeroth-order state. 
Employing a two-body nuclear Hamiltonian built from $\chi$EFT, numerical calculations of the deuteron explored the performance of a hierarchy of mean-field-based methods based upon the concepts of symmetry breaking and restoration.
While too simplistic mean-field approximation schemes do not account for the physics of the deuteron, the d$np$VAPNP+PAV method dubbed in this work as ``mean field on steroids'' was shown to deliver an essentially exact description of the deuteron ground-state. It includes deformation, neutron-proton pairing as well as the restoration of neutron and proton numbers (angular momentum and parity) before (after) variation.
In particular, the resulting binding energy, magnetic dipole moment, electric quadrupole moment and point-proton root-mean-square radius of the deuteron $J^\pi=1^+$ ground-state agree with the reference values obtained from NCSM calculations with sub-percent accuracy. 
By further putting the system into a harmonic trap, the method was also shown to provide nearly perfect values of the $np$ scattering length and effective range in the ${}^{3}S_1$ channel.

Two key ingredients happened to be crucial to reach such a result: i) the inclusion of neutron-proton pairing through the mixing of proton and neutron single-particle states in the Bogoliubov reference state and ii) the energy minimization in presence of proton- and neutron-number projection. Point i) can be connected with previous works dedicated to low-density symmetric nuclear matter where the $np$BCS method was shown to become equivalent to solving the two-body Schrödinger equation for the deuteron \cite{Baldo1995a,Lomabardo2001a}.  Contrary to infinite matter though, the description of a finite nucleus makes it important to work with a wave function containing the exact neutron and proton numbers, i.e.\ to work within the particle-number-restoration formalism.
In the region of very light nuclei where the energy is rapidly changing with the number of nucleons as well as as where proton- and neutron-number fluctuations of the Bogoliubov reference state are of the same order of magnitude as their average values, 
it becomes mandatory to include the particle-number projection before variation, which explains point ii).

The fact that the deuteron can be exactly described at the level of a ``boosted'' mean-field level, i.e.\ at $\basissize^4$ cost (in fact closer to $\basissize^{3.4}$), is counter-intuitive but can actually be supported in several ways, e.g.\ it is known that the two-body system can be solved exactly at (essentially) $\basissize^4$ cost via coupled cluster at the singles and doubles level. 
Therefore, the deuteron constitutes a unique system (so far) where a diagonalization method (NCSM), a ``vertical'' expansion method (CCSD) and a ``horizontal'' expansion method (MFS, this work) all deliver the same exact result.
In addition, obtaining an exact description while working with a low-dimensional linear combination of non-orthogonal product states is of interest when moving to heavier open-shell nuclei. Indeed, this class of states, whose ultimate representatives are states at play in the projected generator coordinate method, can capture strong static correlations all the way to heavy systems while following the $\basissize^4$ scaling~\cite{Bally21a}.
When moving up in mass, it becomes necessary to resum dynamical correlations on top of such reference states via, e.g., a perturbative expansion~\cite{Frosini22a,Frosini22c}. 
It is thus of interest to characterize the quantitative need to go beyond a boosted mean-field level as $A$ increases in order to remain at the sub-percent accuracy level. In this context, one objective in the near future is to extend the present analysis to nuclei heavier than the deuteron where three-nucleon forces start to operate, i.e.\ $^{3}$H and $^{3,4}$He.

A second interesting perspective of the present work concerns the question of the renormalizability of $\cancel{\pi}$EFT calculations at leading order~\cite{Kievsky:2018xsl,Schiavilla:2021dun} given that many-body solutions are bound to be inexact\footnote{Notice that the renormalizability of LO calculations can not only be compromised by the inexact nature of the solution to the A-body Schrödinger equation but also by the incomplete character of the Hamiltonian employed, i.e.\ by the necessity to include many-body forces beyond three-body ones as the nuclear mass increases.} as the nuclear mass increases~\cite{Drissi2020a}. Traditionally, the renormalizability of the two-body part of the leading-order $\cancel{\pi}$EFT Hamiltonian is set up on the basis of an exact solution of the two-body system. Thus, any candidate many-body approximation must deliver an exact solution of the two-body system in order to have a chance to deliver renormalizable predictions.  Identifying a zeroth-order (boosted) mean-field state satisfying renormalization invariance at $\basissize^4$ cost might simplify the task of fulfilling it once dynamical corrections are included on top of it. Our goal is thus to repeat the present study based on the $\cancel{\pi}$EFT LO Hamiltonian while varying the regularization cutoff over a wide interval in order to check the renormalizability of the results. In practice, this will actually constitute a challenging task requiring specific developments given that large cutoffs effectively require the use of very large basis sizes.

%=======================================================================
%=======================================================================
%=======================================================================
\begin{acknowledgements}
We would like to thank G.~Hagen for helping us benchmark our numerical implementation of the center-of-mass Hamiltonian and for providing us with CCSD results for the deuteron, T.~Miyagi for providing us with the NCSM results, as well as M.~Bagnarol for useful discussions.
This work was performed using HPC resources from GENCI-TGCC (Contract No.\ A0150513012) and CCRT (TOPAZE supercomputer) at Bruyères-le-Châtel.
The work of A.S.~was supported by the European Union’s Horizon 2020 research and innovation program under grant agreement No 800945 -- NUMERICS -- H2020-MSCA-COFUND-2017.
\end{acknowledgements}
%
%=======================================================================
%
%\bibliographystyle{apsrev4-2}
\bibliographystyle{apsrev}
%\bibliographystyle{unsrtnat}

%\begingroup
\bibliography{biblio}
%\endgroup
%
%===================================================================
%
\end{document}